\documentstyle[aps,prb,psfig,floats]{revtex}
 
\begin{document}
\title{Transfer-Matrix Monte Carlo Estimates of Critical Points 
in the Simple Cubic Ising, Planar and Heisenberg Models}

\author{M.P. Nightingale,}
\address{Department of Physics, University of Rhode Island, Kingston,
RI 02881}
\author{H.W.J. Bl\"ote,}
\address{Faculty of Applied Physics, Delft University of Technology,
P.O. Box 5046, 2600 GA Delft, The Netherlands}

\maketitle

\begin{abstract}
The principle and the efficiency of the Monte Carlo transfer-matrix
algorithm are discussed.  Enhancements of this algorithm are
illustrated by applications to several phase transitions in lattice
spin models. We demonstrate how the statistical noise can be reduced
considerably by a similarity transformation of the transfer matrix
using a variational estimate of its leading eigenvector, in analogy
with a common practice in various quantum Monte Carlo techniques.  Here
we take the two-dimensional coupled $XY$-Ising model as an example.
Furthermore, we calculate interface free energies of finite
three-dimensional O($n$) models, for the three cases $n=1$, 2 and 3.
Application of finite-size scaling to the numerical results yields
estimates of the critical points of these three models. The statistical
precision of the estimates is satisfactory for the modest amount of
computer time spent.
\end{abstract}
\pacs{02.70.Lq,05.70.Jk,75.10.Hk,05.70.Jk,64.60.Fr}

\section{Introduction}\label{sec.introduction}

Many important problems in computational physics and chemistry can be
reduced to the computation of dominant eigenvalues of matrices of high
or infinite order.  Among the numerous examples of such matrices are
quantum mechanical Hamiltonians and transfer matrices.  The latter were
introduced in statistical mechanics by Kramers and Wannier in 1941 to
study the two-dimensional Ising model \cite{KraWa}, and ever since,
important work on lattice models in classical statistical mechanic has
been done with transfer matrices, producing both exact and numerical
results.

The analogy of the time-evolution operator in quantum mechanics and the
transfer matrix in statistical mechanics allows the two fields to share
numerous techniques.  Specifically, a transfer matrix ${\rm\bf T}$ of a
statistical mechanical lattice system in $d$ dimensions often can be
interpreted as the evolution operator in discrete, imaginary time $t$
of a quantum mechanical analog, as is well known.  That is, ${\rm\bf
T}\approx\exp(-t{\cal H}),$ where $\cal H$ is the hamiltonian of a
system in $d-1$ dimensions, the quantum mechanical analog of the
statistical mechanical system.  From this point of view, the
computation of the partition function and of the ground-state energy
are essentially the same problems: finding the largest eigenvalue of
${\rm\bf T}$ and of $\exp(-t{\cal H})$, respectively.

The transfer-matrix Monte Carlo method used in this paper employs an
algorithm as simple as the diffusion Monte Carlo algorithm, which was
developed to compute the dominant eigenvalue of the evolution operator
$\exp(-t{\cal H})$ .  In contrast to diffusion Monte Carlo,
transfer-matrix Monte Carlo provides exact eigenvalues, subject only to
statistical noise and as qualified below in
Section~\ref{sec.basic.algorithm}.  More specifically, unlike
transfer-matrix Monte Carlo, diffusion Monte Carlo suffers from a
systematic error, the time-step error, because of the necessity to
employ an approximate, short-time evolution operator.  Similar errors
are also found in path-integral Monte Carlo and, in general, in all
approaches based on the Trotter formula \cite{Trot}.  An alternative,
related approach, {\it viz.} Green function Monte Carlo, used to
compute the dominant eigenvalue of $({\cal H} -E)^{-1}$, where $E$ is
close to the ground state energy, does not suffer from a time-step
error, and, from that point of view, Green function Monte Carlo is more
elegant than diffusion Monte Carlo. However, the Green function Monte
Carlo algorithm is considerably more complicated, and enhancement of
that algorithm by the variance reduction techniques discussed below,
has its limitations.

From an orthodox complexity theory point of view, {\em exact} numerical
transfer-matrix computations for lattices in more than one dimension
are intractable, since the order of transfer matrices grows
exponentially with the number of lattice sites in a transfer slice.
Standard Monte Carlo methods in statistical mechanics, on the other
hand, statistically sample the Boltzmann distribution, typically
employing some variant of the Metropolis algorithm.  One can argue that
Monte Carlo methods are of polynomial complexity in the system size, at
least for certain important physical observables. This raises the
question of the ultimate utility of the transfer matrix for
computational purposes.

In many cases, one is interested in the behavior of systems in the
thermodynamic limit.  For critical systems in particular, one has to
rely on finite-size scaling and extrapolation methods to extract the
relevant information from the computations.  The transfer-matrix method
has advantages in both respects.  Firstly, one can compute the spectrum
of the transfer-matrix method virtually to machine precision, which
permits extrapolation without serious loss of numerical accuracy.
Secondly, a large body of numerical evidence suggests that the
transfer-matrix spectrum has weaker corrections-to-scaling than
quantities commonly computed by standard Monte Carlo. Clearly, also the
transfer-matrix Monte Carlo method takes advantage of the weakness of
the corrections-to-scaling. Unfortunately, statistical noise is
introduced, but this can be substantially reduced by the use of
optimized trial eigenvectors, by virtue of which the Monte Carlo
process is in effect only used to compute {\em corrections} to an
already sophisticated approximation.

If one could neglect the correlations introduced by the re-weighting
step of the transfer-matrix Monte Carlo algorithm [see the split/join
steps (\ref{step2.1}) and (\ref{step2.2}) in the algorithm given in
Section~\ref{sec.MCimplem}] and if one could ignore the resulting loss
of efficiency of the transfer-matrix Monte Carlo algorithm, this method
would be a solution to the exponential growth problem mentioned above
\cite{Runge.92.b}.  In addition, transfer-matrix Monte Carlo would be
completely free of critical slowing down, since the correlation {\em
time} of the algorithm is equal to the correlation {\em length} of the
slices used in the definition of the transfer matrix. Again, the use of
optimized trial eigenvectors can serve to reduce the detrimental effect
of the multiplicative re-weighting.

Another feature of the Monte Carlo transfer matrix, which can
contribute to a reduction of the correlation time of the stochastic
process, is that moves are effectively made at surface sites. This
makes it much easier to overcome the barriers some systems present to
standard Monte Carlo algorithms.  An example of such a system is the
$XY$-Ising model discussed in Ref.~\onlinecite{NighGraKos95}.

The layout of this paper is as follows.  In
Section~\ref{sec.basic.algorithm} we review the basic Monte Carlo
algorithm to determine transfer-matrix eigenvalues by means of a
statistical implementation of the power method.  Apart from relatively
minor details, the algorithm given in Section~\ref{sec.basic.algorithm}
is the same as the one discussed in
Refs.\onlinecite{NB.PRL.60,TMreview,UNR}.  Section~\ref{sec.var.red}
describes the similarity transformation of the transfer matrix, which
leads to a pronounced decrease of the statistical errors of the Monte
Carlo process.  Section~\ref{sec.var.red} in particular describes in
detail the construction of a variational approximation of the
eigenstate associated with the largest eigenvalue.  This approximate
eigenstate yields the similarity transformation used to reduce the
statistical errors of the algorithm.  Details of the speed-up of the
algorithm are presented at the end of Section~\ref{sec.var.red} a
coupled $XY$-Ising model in two dimensions.  Finally,
Section~\ref{sec.applic} contains applications of the transfer-matrix
Monte Carlo method to three-dimensional O($n$) models for $n=1$, 2 and
3.  Preliminary discussions of the the work discussed in
Sections~\ref{sec.var.red} and \ref{sec.applic} were published
elsewhere \cite{TMreview,NighGraKos95}.

\section{Monte Carlo implementation of the power
method}\label{sec.MCimplem} \label{sec.basic.algorithm}

Consider an operator ${\rm\bf T}$ of which we want to compute the
dominant eigenvalue.  Let  ${\rm\bf T}$ be represented by matrix
elements $\langle R|{\rm\bf T} | S \rangle =T_{RS}$, where $| R
\rangle$ and $| S \rangle$ are basis states of the physical system
under consideration.  These states will be treated here as discrete.
For Monte Carlo calculations, the distinction between continuous and
discrete states is a minor technicality; in the discussion below,
generalization to the continuous case follows immediately by replacing
the appropriate sums by integrals and replacing Kronecker by Dirac
$\delta$-functions.

Perhaps the simplest way to calculate the dominant eigenvalue of a
matrix or integral kernel is the power method.  That is, choose an
arbitrary initial state $|u^{({0})}\rangle$ and compute iteratively:
\begin{equation} |u^{({t+1})}\rangle = {1 \over c_{t+1}} {\rm\bf T}
|u^{({t})}\rangle, \label{power.method} \end{equation} where $c_{t+1}$
is a constant chosen so that $|u^{({t+1})}\rangle$ is normalized or in
some other convenient standard form.  For $t\rightarrow\infty$, the
constants $c_t$ approximate the dominant eigenvalue $\lambda_0$ of
{\rm\bf T}\, and the vectors $|u^{({t})}\rangle$ converge to the
corresponding eigenvector.

To implement Eq.~(\ref{power.method}) by Monte Carlo,
$|u^{({t})}\rangle$ is represented by a sequence of $N_t$ walkers.
Each of these walkers is a pair $(R_\alpha,w_\alpha),\ \alpha=1,\dots
,N_t$. The variable $R_\alpha$ of a walker represents a possible
configuration of the system described by  ${\rm\bf T}$, and $w_\alpha$
represents its statistical weight. The latter quantity is subject to
the condition $w_{\rm l}<w_\alpha<w_{\rm u}$, where $w_{\rm l}\ {\rm
and}\ w_{\rm u}$ are bounds introduced so as to keep all weights
$w_\alpha$ of the same order of magnitude, which improves the
efficiency of the algorithm.  This sequence of walkers represents a
(sparse) vector with components
\begin{equation}
{\underline u}^{({t})}_{R} = \sum_{\alpha=1}^{N_t} w_\alpha 
\delta_{R,R_{\alpha}},
\label{uc.walkers}
\end{equation}
where $\delta$ is the usual Kronecker $\delta$-function.  The
underscore is used to indicate that the ${\underline u}^{({t})}_{R}$
represent a stochastic vector $|{\underline u}^{({t})}\rangle$.  A
stochastic process will be defined presently with transition
probabilities such that $c_{t+1}|{\underline u}^{({t+1})}\rangle $ has
a conditional expectation value equal to $ {\rm\bf T} |{\underline
u}^{({t})}\rangle$ for any given sequence of walkers representing
$|{\underline u}^{({t})}\rangle$.  In practice, one has to average over
the stationary state of a stochastic process in which the constants
$c_t$ are determined on the fly, so that $c_{t+1}$ and $|{\underline
u}^{({t+1})}\rangle$ are correlated.  As a consequence, there is no
guarantee that the stationary state expectation value of $|{\underline
u}^{({t})}\rangle$ is {\em precisely} an eigenstate of ${\rm\bf T}$, at
least not for finite $N_t$.  The same mathematical problem occurs if
one takes the time-average of Eq.~(\ref{power.method}) in the presence
of noise correlated to the $c_t$.  The resulting bias
\cite{Hetherington,NB.PRB.33} has also been discussed in the context of
diffusion Monte Carlo \cite{UNR}. 

To define the stochastic process, Eq.~(\ref{power.method}) is rewritten
as
\begin{equation}
u^{({t+1})}_{R}={1 \over c_{t+1}}\sum_S P_{RS} D_S u^{(t)} S, 
\label{factor.tm}
\end{equation}
where
\begin{equation}
D_S=\sum_R T_{RS}$ and $P_{RS}=T_{RS}/D_S.
\label{DP}
\end{equation}

Eq.~(\ref{factor.tm}) describes a process represented by a Monte Carlo
run which, in addition to a few initial equilibration sweeps, consists 
of a time series of a little over $M_0$ sweeps over all walkers at times
labeled by $t=\ldots,0,1,\ldots,M_0$.  The sweep at time $t$ consists
of two steps designed to perform stochastically the matrix
multiplications in Eq.~(\ref{factor.tm}).  Following Nightingale and
Bl\"ote \cite{NB.PRB.33}, the process is defined by the following
steps, which transform the generation of walkers at time $t$ into the
the generation at time $t+1$.  Variables pertaining to times $t$ and
$t+1$ will be denoted respectively by unprimed and primed symbols.
\begin{enumerate}
\item \label{step1}
Update the old walker $(S_{\alpha},w_{\alpha})$ to yield a temporary
walker $(S'_{\alpha},w'_{\alpha})$ according to the transition
probability $P_{S'_{\alpha}S_{\alpha}}$, where
$w'_\alpha=D_{S_\alpha}w_\alpha/c'$, for $\alpha =1,\ldots ,N_t$.  The
next step can change the number of walkers. To maintain their number
close to a target number, say $N_0$, choose $c' =\hat
\lambda_0(N_t/N_0)^{1/s}$, where $\hat \lambda_0$ is a running estimate
of the eigenvalue $\lambda_0$ to be calculated, where $s\ge 1$ (see
below).
\item \label{step2}
From the temporary walkers construct the new generation of walkers
as follows:
\begin{enumerate}
\item \label{step2.1}
Split each walker $(S',w')$ for which $w'>b_{\/\rm u}$ into two walkers
$(S',{1\over2}w')$.  The choice $b_{\/\rm u}=2$ is a reasonable one.
\item \label{step2.2}
Join pairs $(R'_{\alpha},w'_{\alpha})$ and $(R'_{\beta},w'_{\beta})$
with $w'_{\alpha}<b_{\/\rm l}$ and $w'_{\beta}<b_{\/\rm l}$ to produce
a single walker $(R'_{\gamma},w'_{\alpha}+w'_{\beta})$, where
$R'_{\gamma}=R'_{\alpha}$ or $R'_{\gamma}=S'_{\beta}$ with relative
probabilities $w'_{\alpha}$ and $w'_{\beta}$.  We chose $b_{\/\rm l}=1/2$.
\item \label{step2.3}
Any temporary walker left single in step (\ref{step2.2}), or for which
$b_{\/\rm l}<w'_{\alpha}<b_{\/\rm u}$, becomes a permanent member of
the new generation of walkers.
\end{enumerate}
\end{enumerate}

The algorithm described above was constructed so that for any given
realization of $|{\underline u}^{({t})}\rangle$, the expectation value
of $c_{t+1} |{\underline u}^{({t+1})}\rangle$, in accordance with
Eq.~(\ref{power.method}), satisfies
\begin{equation}
{\rm E} \left( c_{t+1} |{\underline u}^{({t+1})}\rangle\right)=
{\rm\bf T} |{\underline u}^{({t})}\rangle,
\label{power.1}
\end{equation}
where ${\rm E}(\cdot)$ denotes the conditional average over the
transitions defined by the above stochastic process.  More generally by
$p$-fold iteration one finds \cite{NB.PRL.60}:
\begin{equation}
{\rm E} \left(\left[\prod^p_{b=1} c_{t+b} \right] |{\underline
u}^{({t+p})}\rangle\right)= {\rm\bf T}^p |{\underline
u}^{({t})}\rangle, \label{power.p}
\end{equation}

The stationary state average of $|u^{({t})}\rangle$ is close to the
dominant eigenvector of ${\rm\bf T}$, but, as mentioned above, it has a
systematic bias when the number $N_t$ of walkers is finite. For
increasing $p$, components of non-dominant eigenvectors can be
projected out and thus the bias is reduced, in principle.
Unfortunately, the variance of the corresponding estimators increases
as their bias decreases.  The reader is referred to Refs.
\onlinecite{Hetherington,NB.PRL.60,TMreview,Runge.92.b} for a more
detailed discussions of this problem.  Suffice it to mention here,
firstly, that $s$ is the expected number of time steps it takes to
restore the number of walkers to its target value $N_0$ and, secondly,
that strong population control ($s=1$) tends to introduce a stronger
bias than weaker control ($s>1$) \cite{UmrigarCeperleyPriv}.

With Eq.~(\ref{power.p}) one constructs an estimator \cite{NB.PRL.60}
of the dominant eigenvector $|u^{({\infty})}\rangle$ of the matrix
{\rm\bf T}:

\begin{equation}
|\hat u^{({p})}\rangle= {1\over M_0}\sum^{M_0}_{t=1}\left(
\prod^{p-1}_{b=0} c_{t-b} \right) |{\underline u}^{({t})}\rangle.
\label{u.estimator}
\end{equation}

More practically, suppose that $\langle\psi_{\rm T} |$ is an
approximate leading eigenbra $\langle\psi_{\rm T} |$ of ${\rm\bf T}$,
and that ${\rm \bf O}$ is an arbitrary operator.  The mixed expectation
value of ${\rm \bf O}$  can be approximated as
\begin{equation}
{\langle\psi_{\rm T} | {\rm \bf O} |u^{({\infty})}\rangle \over
\langle\psi_{\rm T} |u^{({\infty})}\rangle} \approx {\langle\psi_{\rm
T} |{\rm \bf O} |\hat u^{({p})}\rangle \over \langle\psi_{\rm T} |\hat
u^{({p})}\rangle}.  \label{mixed.estimator}
\end{equation}

An important special case is obtained by choosing in this expression
${\rm \bf O}={\rm\bf T}$ and $\langle\psi_{\rm T} |R\rangle =1$ for all
$R$.  The latter corresponds to the infinite-temperature approximation
for the trial state and in that case, Eq.~(\ref{mixed.estimator})
reduces to an estimator for the dominant eigenvalue of ${\rm\bf T}$:
\begin{equation}
\lambda_0 \approx {\sum_{t=1}^{M_0} \left( \prod^{p}_{b=0}c_{t-b} 
\right) W^{(t)} \over \sum_{t=1}^{M_0} \left( \prod^{p-1}_{b=0}c_{t-b} 
\right) W^{(t-1)}},
\label{growth.estimator}
\end{equation}
where
\begin{equation}
W_t=\langle\psi_{\rm T} |{\underline
u}^{({t})}\rangle=\sum_{\alpha=1}^{n_t}w_\alpha^{(t)}.
\end{equation}

For the above special choice of the trial bra $\langle\psi_{\rm T} |$,
Eq.~(\ref{mixed.estimator}) becomes the expression for the surface
expectation value of ${\rm \bf O}$ in the geometry shown on the right
in Fig.~\ref{lreivecs}.  Although we have used the transfer-matrix
algorithm only for the computation of the dominant eigenvalue of the
transfer matrix for the applications discussed in this paper, it should
be mentioned for completeness that one can also compute bulk
expectation values, at least asymptotically, as follows.

\begin{figure}
\centerline{\psfig{figure=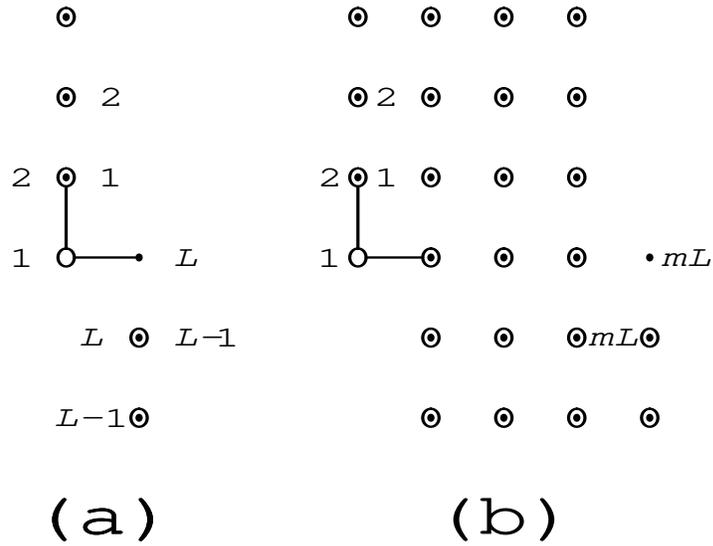,height=4.2truein,width=4.2in}}
\caption[c1]{\footnotesize \narrower
Illustration of the calculation of correlation functions involving
spins in the bulk below the surface layer. 
Site labels before the addition of the new spin (open circle) appear 
to the right, and the new labels to the left of a lattice point.}
\label{bulcor}
\end{figure}

One can represent the Kramers-Wannier transfer matrix by the graph
shown in Fig.~\ref{bulcor}.a.  This matrix transfers from an old slice
to a new one, with slices represented respectively by small full and
large open circles.  The process adds only one new site: the open
circle labeled 1. One site, the small closed circle labeled $L$, is
about to disappear into the bulk.  Coincidences of both types of
circles represent Kronecker-$\delta$ functions in the transfer matrix
[see Eq.~(\ref{tm.ising})].  The solid lines stand for interactions
added in one transfer operation.  One can define a transfer matrix with
extended slices consisting of $m$ of the original, minimal slices.  The
dominant eigenvector of this extended transfer matrix is simply the
original eigenvector multiplied by the Boltzmann weight associated with
the portion of the lattice containing variables that have not yet been
summed over.  Eq.~(\ref{mixed.estimator}), used with any operator in
which occur only variables of slice $m$, becomes a bulk expectation
value for $m \to \infty$.  The implementation of this concept is called
{\it forward walking} in the context of quantum Monte Carlo
\cite{Kalospri,Runge.92.a}, and this only requires extending the
walkers so that their states correspond to the extended slices
introduced above.  This increases the memory requirements and the cost
of splitting a walker, but otherwise the efficiency of the algorithm is
not affected.

\section{Variance Reduction (Importance Sampling) and Trial Vectors}
\label{sec.var.red}

In principle, if $\langle\psi_{\rm T} |$ equals an exact eigenbra of
the operator $O$ in equation Eq.~(\ref{mixed.estimator}), the
right-hand side of the expression is a zero-variance estimator.  In
general, no exact eigenvectors are known, but even an approximation may
yield a substantial reduction of statistical noise.  A more efficient
well-known \cite{ck.79} way to exploit an approximate left eigenbra
$\langle\psi_{\rm T} |$ to reduce variance works by application of the
method described above to a similarity transform of the original
operator ${\rm\bf T}$.  This transformation is defined by:
\begin{equation}
{\rm\bf \tilde T}={\rm\bf I} {\rm\bf T} {\rm\bf I}^{-1},
\label{Ttilde}\\
\end{equation}
where ${\rm\bf I}$ is diagonal in the configuration presentation, and
is defined as
\begin{equation}
{\rm\bf I}=\sum_R |R\rangle \langle\psi_{\rm T} |R \rangle \langle R |.
\label{I}
\end{equation}

Ideally, $\langle\psi_{\rm T} |$ would equal the exact dominant
eigenbra of ${\rm\bf T}$.  In that case, the stochastic process defined
as above, but with ${\rm\bf T}$ replaced by ${\rm\bf \tilde T}$, would
become optimally efficient and in fact would lack critical slowing
down.  For such an ideal process ${\rm\bf \tilde D}$, defined as in
Eq.~(\ref{DP}) as a function of ${\rm\bf \tilde T}$, would be a
constant times the unit matrix.  The walker weights would no longer
fluctuate so that birth and death process would no longer occur. The
walkers would evolve into a statistically independent ensemble.  The
estimator given in Eq.~(\ref{mixed.estimator}), appropriately
transformed, would have zero variance. The transformed bra
$\langle\tilde\psi_{\rm T}| = \langle\psi_{\rm T}| {\rm\bf I}^{-1}$
would have all elements equal to unity in the configuration
representation. In other words, ${\rm\bf \tilde T}$ would be
represented by a stochastic matrix, which would eliminate re-weighting
of walkers and the concomitant split/join step in the algorithm.

In the absence of {\em exact} eigenbras, approximations may be obtained
by variational methods.  The variational expression for the leading
eigenbra $\langle\psi_{\rm T}|$ can conveniently be cast in the form of
an effective surface Hamiltonian with pair interactions between nearest
neighbors, next-nearest neighbors, and so on. These interactions are
treated as variational parameters and can be determined from an
analysis of the walker population \cite{UWW}.

Since generalization to higher dimensions and models with different
microscopic variables is straightforward, it will suffice to consider
the Kramers-Wannier transfer matrix for the two-dimensional Ising model
to explain the construction of trial vectors used in the applications
discussed in Section \ref{sec.applic}.

\begin{figure}
\centerline{\psfig{figure=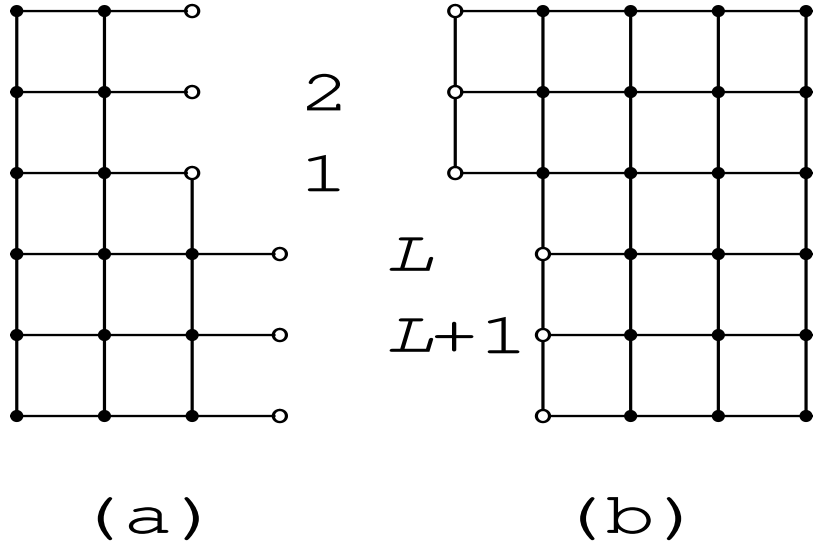,height=4.2truein,width=4.2in,%
rheight=4.2truein,rwidth=4.2truein}}
\caption[c1]{\footnotesize \narrower
Illustration of left and right eigenvectors of the transfer matrix.}
\label{lreivecs}
\end{figure}

For a simple quadratic lattice of $M$ sites, wrapped on a cylinder with
a circumference of $L$ spins and helical boundary conditions, the
transfer matrix for the Ising model is
\begin{equation}
T_{S,R} ={\rm e}^{K(s_1 r_1 + s_1 r_L)}
\prod_{i=1}^{j-1}\delta_{s_i,r_{i+1}},
\label{tm.ising}
\end{equation}
with $S=(s_1,s_2,\dots,s_L)$ and $R\,=(r_1,r_2,\dots,r_L)$, where the
$s_i=\pm 1$ and $r_i\pm 1$.  The conditional partition function of the
lattice of $M$ sites, subject to the restriction that the spins on the
left-hand edge are in state $R$, as illustrated in
Figure~\ref{lreivecs}, is denoted $Z_{M}(R)$.  One has
\begin{equation}
Z_{M+1}(S)=\sum_{R}T_{S,R}Z_{M}(R).
\label{Z}
\end{equation}

Obviously, for $M\to\infty$ the restricted sums $Z_{M} (R)$ are
proportional to the components $u^{({\infty})}_{R}$ of the dominant right
eigenvector of the transfer matrix.  The eigenvector is represented by
the graph on the right in Figure~\ref{lreivecs}. Full circles indicate
spins that have been summed over, while the fixed surface spins are
represented by the open circles; each bond represents a factor $\exp(K
s_i s_j)$.  The left eigenvector, which is the one that has to be
approximated by an optimized trial vector, is represented by the graph
on the left.  In passing, we mention the following relation between left
and right eigenvectors, which follows by inspection of the graphs:
\begin{equation}
\langle {u^{(\infty)}}| S\rangle= \prod_{i=1}^{L-1}{\rm e}^{K s_i
s_{i+1}}\langle {U(S)}{|u^{({\infty})}\rangle}, \label{left.right}
\end{equation}
where $U$ is the reflection operator: $U(S)=(s_L,s_{L-1},\dots,s_1)$.

A similarity transformation of the transfer matrix {\rm\bf T}\ can be
introduced by dividing up the interaction energies between the columns
differently. That is, $h$ is introduced by writing
\begin{equation}
T_{S,R} \equiv {\rm e}^{h(S,R)}.
\end{equation}
A transformation $h \to \tilde h$ is defined by
\begin{eqnarray}
\tilde h(S,R)=g(S) +h(S,R)-g(R),
\label{htilde}\\
\tilde T_{S,R}= \hat\psi_{\rm T}(S)T_{S,R}/\hat\psi_{\rm T}(R),
\label{Tmtilde}\\
\hat\psi_{\rm T}(S)={\rm e}^{g(S)}.
\label{gamma.ising}
\end{eqnarray}

For purposes of variance reduction, versatile trial vectors that
capture some of the essential physics without seriously slowing down
computations, can be chosen of the following form
\begin{equation}
\hat\psi_{\rm T} (S)={\rm e}^{\sum_{i,j}^\ast K_{ij} s_i s_j},
\label{trial.ising}
\end{equation}
a form reminiscent of the Jastrow functions used for quantum many-body
systems.  The asterisk in the sum over pairs indicates that the
$K_{ij}$ are truncated for distances greater than a couple of lattice
spacings.

The couplings $K_{ij}$ in Eq.~(\ref{trial.ising}) are variational
parameters. They can be determined efficiently with the Monte Carlo
scheme introduced by Umrigar, Wilson and Wilkins \cite{UWW}, {\it
i.e.}, by minimization of the variance of $\tilde D(S)$, where the
variance is approximated by a weighted sum over the states of the
walkers of one generation, during the initial stage of the Monte Carlo
run.  This procedure is efficient and stable as long as the $K_{ij}$
are truncated with care, in which case it is perfectly feasible to use
as many as 50 to 100 different parameters.

The magnitude of the $K_{ij}$ is expected to increase with the strength
of the correlations between surface spins.  Since all correlations
between surface spins for the left eigenvector have to be propagated
through the lattice on the left, as illustrated in
Figure~\ref{lreivecs}, one expects that for high temperatures, {\it
i.e.}, small $K$,
\begin{equation}
K_{ij} \propto K^{d_{ij}},
\end{equation}
where $d_{ij}$ is the length of the shortest path along edges connected
by bonds between sites $i$ and $j$.  By inspection of the graph in
Figure~\ref{lreivecs}, we therefore expect the following partial
ordering in decreasing strength of interaction and increasing $d_{ij}$
\begin{equation}
\left.
\begin{array}{l}
d_{1L}=2,\\
d_{12}=d_{23}=...=d_{L-1,L}=d_{1,L-1}=3,\\
d_{13}=d_{24}=...=d_{L-2,L}=d_{1,L-2}=4\\
d_{14}=d_{25}=...=d_{L-3,L}=d_{1,L-3}=d_{2,L}=5\\
\vdots
\end{array}
\right\}
\label{paths}
\end{equation}

It is important to note that if $K_{ij}=K_{i+1,j+1}$ the corresponding
factors cancel in the transformed transfer matrix ${\rm\bf \tilde T}$
for $2 \le i \le L-2$,  since $s_i=t_{i+1}$ for non-vanishing
transfer-matrix elements.  For reasons of efficiency it is therefore
advantageous to have this equality satisfied as often as possible.
Unfortunately, helical boundary conditions introduce a step which
destroys translation symmetry on the surface and renders the partial
ordering in Eq.~(\ref{paths}) insufficient.  For example, sites $1$ and
$2$ are more strongly correlated than $2$ and $3$, and correlations
keep decreasing through pair $(L-1,L)$.  Consequently,
$K_{12}>K_{23}>\dots>K_{L-1,L}$.

In practice, the {\it differences} between the $K_{ij}$ with $d_{ij}=3$
are frequently greater than the higher-order $K_{ij}$.  Then, it is
necessary to treat $K_{12}$ and $K_{23}$ as different parameters of the
trial vector.  An efficient compromise is to treat $K_{ij}$ in which
site $1$ or $L$ participate as different.  The same applies to all
$K_{ij}$ for which the shortest path between $i$ and $j$ straddles the
step on the surface.  To summarize, we distinguish different types of
pairs of sites $(i,j)$ both on the basis of the distance $d_{ij}$ and
to some extent on the location of the pair, enforcing as much
translation invariance as possible.

Clearly, none of the above depends only on lattice geometry or the
Ising nature of the variables.  In general, only a method is required
to generate lists of lattice sites separated by various distances
$d_{ij}$. These can be constructed with simple graph theoretic tools
such as incidence matrices, which makes it possible to deal with
different dimensions and lattice types in an identical fashion once the
pertinent incidence matrix has been defined.

To illustrate the efficiency and flexibility of this technique for
constructing trial vectors, we use the $XY$-Ising model.  It consists
of coupled Ising and planar rotator degrees of freedom on a simple
quadratic lattice. On each lattice site there are two variables:  $s_i
\pm 1$ and ${\rm\bf n_i}$, a two-component unit vector.  The reduced
Hamiltonian ---divided by $-k_{\rm B}T$--- is given by
\begin{equation}
H=\sum_{(i,j)}\left( A\,{\rm\bf n}_i \cdot {\rm\bf n}_j + 
B\, {\rm\bf n}_i \cdot {\rm\bf n}_j s_i s_j+ C s_i s_j\right).
\label{XY.Ising.ham}
\end{equation}

We consider the special case $A=B$ and only from the point of view of
the performance of transfer-matrix Monte Carlo algorithm. For a
discussion of the physics of this model the reader is referred to
Ref.~\onlinecite{NighGraKos95}.  The trial vectors discussed above for
the Ising model have an immediate generalization:
\begin{equation}
\psi_{\rm T} =\exp\left(\sum_{i,j}^\ast 
\left( A_{i,j}\,{\rm\bf n}_i \cdot {\rm\bf n}_j  + 
B_{i,j}\, {\rm\bf n}_i \cdot {\rm\bf n}_j s_i s_j+ C_{i,j} 
s_i s_j\right)\right) .
\label{trial.xy.ising}
\end{equation}

The truncation scheme introduced above for the Ising model is purely
geometrical, and therefore carries over without changes to the
$XY$-Ising model.  It should, however, be noted that there are models
and choices of transfer matrices to which the above scheme is not
applicable.  Ref.~\onlinecite{GraNigh93} contains a discussion and an
example of such a case.

Table~\ref{tab.xy-ising} shows the estimates of the dominant eigenvalue
of $XY$-Ising model for trial vector truncated at different values of
$d_{ij}.$  As can be seen by comparing the first and last lines of the
table, the variance in the estimate of the eigenvalue is reduced by a
factor 300 for a fixed number of Monte Carlo steps.  Taking into
account that the computer time per step doubles, this constitutes a
speed-up by a factor of 150.

\begin{table}
\caption[tc1]{\footnotesize \narrower
Estimated eigenvalue and standard deviations for the $XY$-Ising model.
These data apply to the point $(A=1.005,C=-0.2285)$ [{\it cf.}
Eq.~(\ref{XY.Ising.ham})] on the line where Ising and $XY$ transitions
coincide. Results are shown for various values of $d_{\rm m}$, the path
length of the cutoff in Eq.~(\ref{trial.xy.ising}).  The results are
for a strip of width $L=20$ and were obtained with a target number of
walkers $N_0=10,000$ and $M_0=1,250L$ generations of which an initial
10\% were discarded. The last column shows the computer time in
arbitrary units needed per time step of one walker.
}
\label{tab.xy-ising}
\bigskip
\begin{tabular}{|d|d|d|d|}
$\lambda_0$ & $\sigma$ & $d_{\rm m}$ &$\mu$s\\
\hline   
34.17406   &   0.0071  & 0  & 15\\
34.20875   &   0.0052  & 2  & 15\\
34.21658   &   0.0015  & 3  & 17\\
34.21418   &   0.00083 & 4  & 19\\
34.21384   &   0.00052 & 5  & 21\\
34.21366   &   0.00049 & 6  & 23\\
34.21379   &   0.00041 & 7  & 26\\
\end{tabular}
\end{table}

\section{Applications}
\label{sec.applic}

As an illustration of the transfer-matrix technique we apply the method
to three-dimensional O($n$) models for $n=1$, 2 and 3, {\it i.e.}\, the
Ising, planar and Heisenberg model. In particular the significance of
the results for of the planar and Heisenberg models goes beyond mere
illustrations. These results are sufficiently accurate to be of some
relevance for the location of the critical points.

The O($n$) spins are located on the simple cubic lattice. The
transfer matrix for an $L\times L\times \infty$ system, with helical
boundary conditions and layers of $N=L^2$ sites each, is a
straightforward generalization of Eq.~(\ref{tm.ising}) and reads:
\begin{equation}
T_{{\rm\bf S}, {\rm\bf R}} = \prod^{N-1}_{i=1}\delta_{{\rm\bf
s}_i,{\rm\bf r}_{i+1}} \exp\left[K{\rm\bf s}_1\cdot({\rm\bf
r}_1+{\rm\bf r}_{L}+{\rm\bf r}_N)\right], \label{tm.On.3d}
\end{equation}
where the ${\rm\bf s}_i$ and ${\rm\bf r}_i$ are $n$-component unit
vectors, ${\rm\bf S},=({\rm\bf s}_1,{\rm\bf s}_2,\dots,{\rm\bf s}_{N})$
and ${\rm\bf R},=({\rm\bf r}_1,{\rm\bf r}_2,\dots,{\rm\bf r}_{N})$.

As discussed above, transfer-matrix Monte Carlo is designed to compute
the dominant eigenvalue $\lambda_0$ of the transfer matrix.  The
reduced free energy per site is $f=-\ln \lambda_0$.  From the free
energy one can calculate the surface tension as the difference in free
energy of two systems: one with ferromagnetic and the other with
antiferromagnetic interactions, if the dimensions are chosen so as to
force an interface in the antiferromagnetic system.  For $L\times
L\times\infty$ systems with helical boundary conditions, to which the
present calculations are restricted, this means that $L$ has to be
even.

Renormalization group theory predicts that the values of $\Delta$, the
reduced interface free energy per lattice site, as a function of
coupling $K$ and system sizes $L$ collapse onto a single curve, at
least close to the critical point $K_{\/\rm c}$ and for sufficiently
big systems.  In terms of the non-linear thermal scaling field
\begin{equation}
 u(K)=K-K_{\/\rm
c}+a(K-K_{\/\rm c})^2+\dots, \label{scalingfield} 
\end{equation}
 this curve
$\Sigma(x)$ is determined by 
\begin{equation} \Delta (u,L) = L^{1-d} \Sigma
(L^{y_{\/\rm T}}u), \label{Deltahomog}
\end{equation}
for a $d$-dimensional
system with a thermal scaling exponent $y_{\/\rm T}$. The function
$\Sigma$ can be expanded in a series:  
\begin{equation}
\Sigma(x)=\sum_{l=0}^{\infty}\Sigma_l x^l, 
\end{equation}
and for O($n$) models behaves for large $x$ as:
\begin{equation}
\Sigma(x)=A_{\Sigma}
x^{[d-1-p(n)]/y_{\/\rm T}}, 
\label{asigma} 
\end{equation} 
where $p(1)=0$ and $p(2)=p(3)=1$.

Eqs. (\ref{Deltahomog}) to (\ref{asigma}) are useful for the
interpretation of the O($n$) transfer-matrix Monte Carlo results for
the interface free energy. These results were obtained using finite
sizes up to $L=12$, and populations typically consisting of 2500 or
5000 walkers. Typical run lengths are 5000 steps, where each step means
the addition of a surface layer of $L \times L$ spins.
Variance-reducing trial vectors [see Eq. (\ref{trial.ising})] were
constructed for path lengths up to 5. As before, the variance of the
Monte Carlo process was observed to decrease considerably with
increasing path length.  For each system size, interface free energies
were obtained for approximately 10 different couplings in a range of
about 10\% around the critical points of the Ising and planar models,
and about 1\% for the case of the Heisenberg model.

\begin{figure}
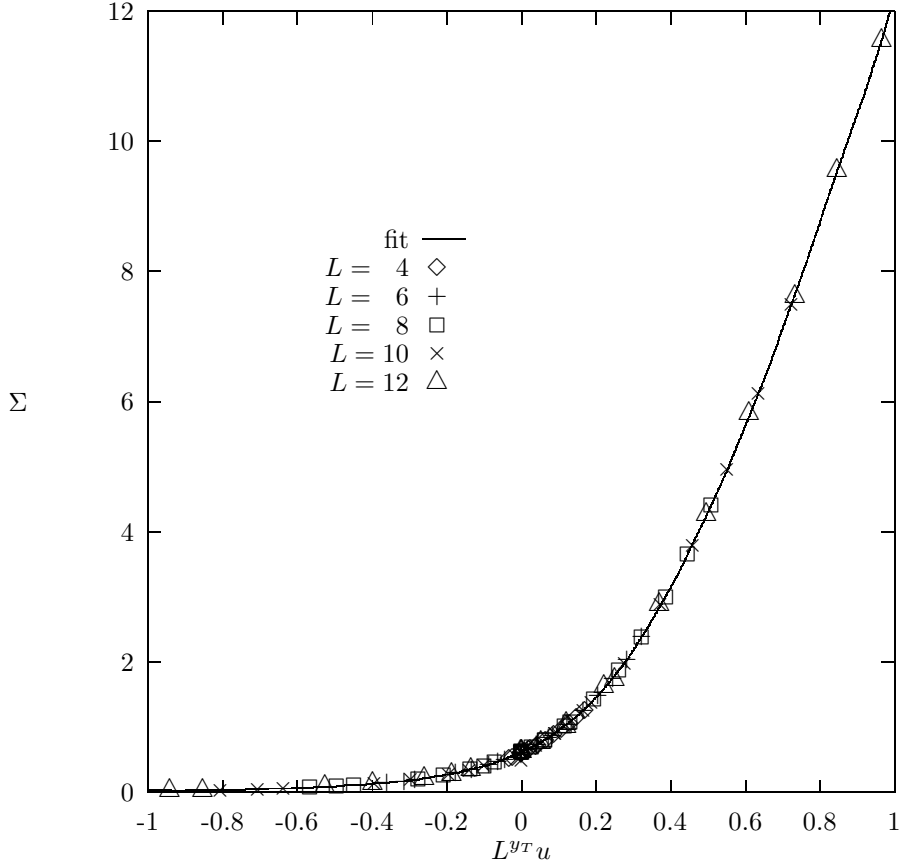
 \centerline{\input O1ScalingPlot}
\caption[is]{\footnotesize \narrower Finite-size scaling plot for the
interface free energy of the three dimensional Ising model.}
\label{scaling.plot.n1} \end{figure}

On the basis of these results for the Ising ($n=1$) case, the function
$\Sigma$ is shown in Figure~\ref{scaling.plot.n1}.  This data collapse
is achieved by means of a least-squares fit with parameters $K_{\/\rm
c},\  y_{\/\rm T},\ a,\ \rm$ and 13 Taylor coefficients $\Sigma_l$, a
generalization of a technique used in the past \cite{NB.PRL.60}.

To check if the system sizes were in the asymptotic finite-size scaling
regime, fits were done both with and without the $6\times 6\times
\infty$ data. The results of these fits are displayed in
Table~\ref{tab.Isingfit} in Appendix A.  To summarize, the results
are:  $K_{\/\rm c}=0.22162\pm 0.00002$ and $y_T=1.584\pm 0.004$ using
data with $L=6$ through $12$; and $K_{\/\rm c}=0.22167\pm 0.00004$ and
$y_T=1.584\pm 0.014$ if the $L=6$ are omitted.  These results agree
well with accurate determinations using other methods (see e.g.
Ref.~\onlinecite{BLH,BHHSM,GT} and references therein) which appear to
cluster about $K_c=0.221655$  (with a margin of about $10^{-6}$) and
$y_T=1.586$ (with a precision of a few times $10^{-3}$).

It is remarkable that the corrections to scaling appear to be very
small, as appears from the data shown in Figure~\ref{scaling.plot.n1}.
In standard Monte Carlo analyses \cite{FerLan} of $L \times L \times L$
systems these corrections are quite prominent, and form an obstacle to
the accurate determination of critical parameters.

The scaling plot shown in Figure~\ref{scaling.plot.n1}, can be used to
determine the amplitude $A_{\Sigma}$ graphically: on a double
logarithmic plot the asymptotic slope of the curve follows from the
known value of the thermal exponent $y_{\/\rm T}$, {\it cf.}
Eq.~(\ref{asigma}). The problem of calculating this amplitude has
attracted considerable attention lately and the reader is referred to a
paper by Shaw and Fisher \cite{ShawFisher} for details and further
references to the literature.  For the largest values of the scaled
temperature variable $x$, we find $A_{\Sigma}'=A_{\Sigma} K_{\/\rm
c}^{2/y_{\rm T}}=1.8$, while the trend with $x$ is an increasing one.
This value is somewhat larger than Mon's \cite{KKMonPRL88} estimate
$A_{\Sigma}=1.58\pm 0.05$, but still in the range $1.4 \le A_{\Sigma}
\le 2.0$ obtained by Shaw and Fisher. As a final comment we note that
Mon's method requires systems of linear dimensions in excess of 48 to
reach the asymptotic infinite-size regime, with an increasing trend of
the estimates of $A_{\Sigma}$ with increasing $x=L^{y_{\rm T}}u$.

A similar analysis was performed for the planar model ($n=2$).  In
comparison with the Ising case, the scaling function $\Sigma$ behaves
more smoothly as a function of $x$, so that a satisfactory fit could be
obtained with fewer Taylor coefficients. The fitted parameters, which
are $K_{\/\rm c},\  y_{\/\rm T},\ a,\ \rm$ and 8 Taylor coefficients
$\Sigma_n$, are shown in Table~\ref{tab.XYfit} of Appendix A.  Our
results for the critical point are $K_{\/\rm c}=0.45410\pm 0.00003$ for
system sizes $L=6$ to 12, and  $K_{\/\rm c}=0.45413 \pm 0.00005$ for
$L=8$ to 12. These values are close to results from series expansions
\cite{Ferer,Ohno} $K_{\/\rm c}=0.45386$ and standard Monte Carlo
calculations \cite{LanPanB} $K_{\/\rm c}=0.4531$ (no errors quoted).
Also our results for the temperature exponent, namely $y_{\/\rm
T}=1.491 \pm 0.003 $ for $L\geq 6$ and $y_{\/\rm T}=1.487 \pm 0.006 $
for $L\geq 8$ are in a good agreement with existing results; we quote
the coupling-constant-expansion value \cite{LGZJ} $y_{\/\rm T}=1.495
\pm 0.005$.

\begin{figure}
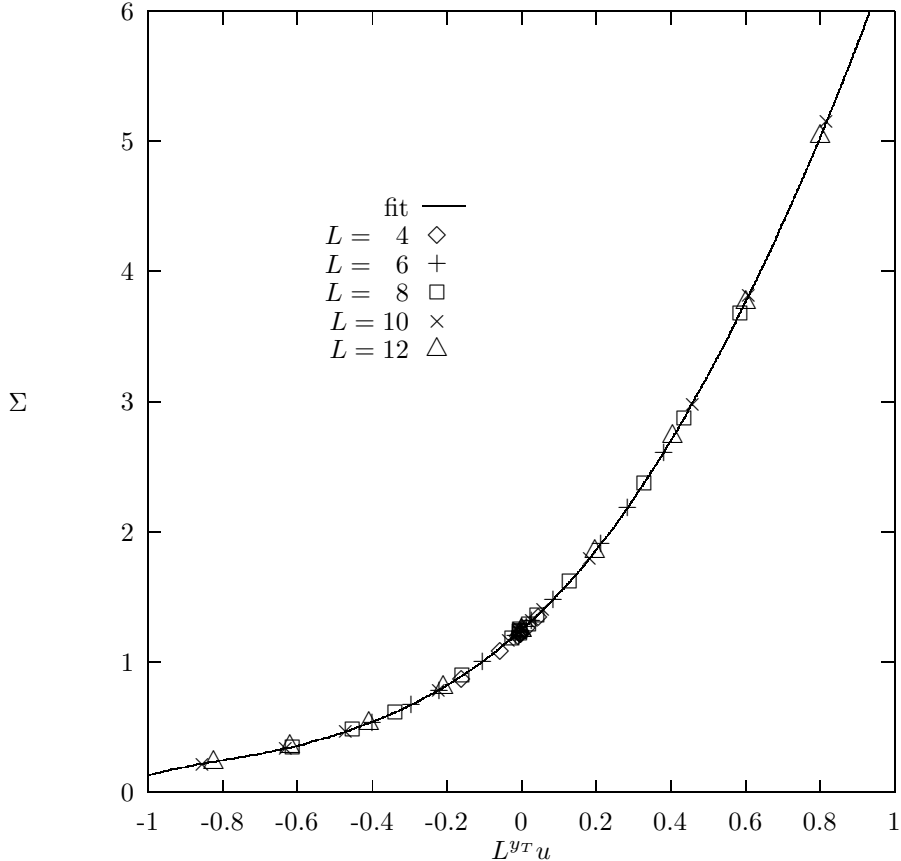

\centerline{\input O2ScalingPlot}
\caption[xy]{\footnotesize \narrower
Finite-size scaling plot for the interface free energy 
of the three dimensional planar model.}
\label{scaling.plot.n2}
\end{figure}
 
Fitted with these parameters the data collapse very well onto
the function $\Sigma$, as shown in Figure~\ref{scaling.plot.n2}. 
Again, this scaling plot can be used to determine the amplitude 
$A_{\Sigma}$ graphically: in this case the asymptotic power-law
exponent is $1/y_{\/\rm T}$. A fit of the data at the highest 
available values of $x=L^{y_{\/\rm T}} u$ leads to $A_{\Sigma}=5.9$, 
while the trend is still increasing with $x$. 

The calculations for the Heisenberg case $n=3$ were clustered in a 
narrow interval around the critical temperature, and were not aimed
at an accurate determination $y_{\/\rm T}$. Thus, the transfer-matrix 
Monte Carlo data could be analyzed by means of a least-square fit with 
less parameters: $K_{\/\rm c}$, $y_{\/\rm T}$ and 3 Taylor coefficients 
$\Sigma_n$. The fit is shown in Table~\ref{tab.Heisfit} in Appendix A.   
The result for $y_{\/\rm T}$ is well within the statistical accuracy, 
equal to the known coupling-constant-expansion value \cite{LGZJ}
$y_{\/\rm T}=1.418$. Including the latter value as a known variable 
in the fits leaves our results for the critical point practically 
unchanged. These are: $K_{\/\rm c}=0.69291\pm 0.00004$
for system sizes $L=6$ to 12, and  $K_{\/\rm c}=0.69294 \pm 0.00008$ 
for $L=8$ to 12. These values are close to results from series 
expansions \cite{RitFis}: $K_{\/\rm c}=0.6916$, and more recently
\cite{Ohno}: $K_{\/\rm c}=0.69294$; and from Monte Carlo 
calculations \cite{ChenFerLan}: $K_{\/\rm c}=0.693035 \pm 0.000037$. 
The difference with our result with the $L=6$ data included could be
interpreted as an indication of a small finite-size effect. 

\begin{figure}
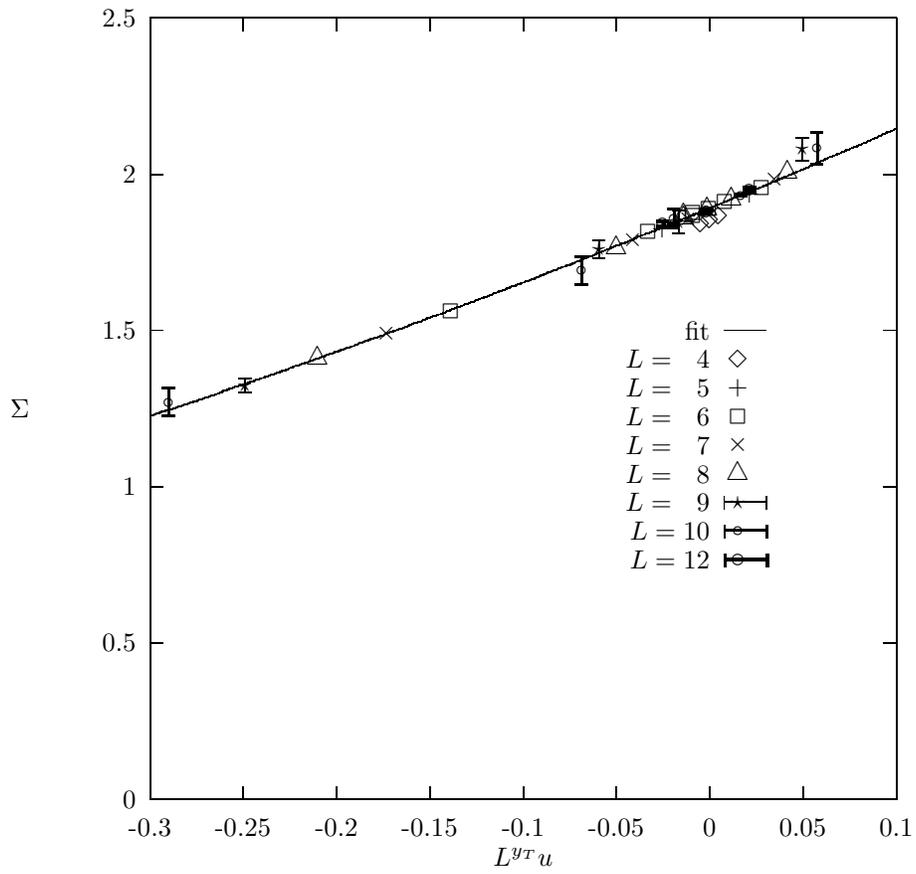

\centerline{\input O3ScalingPlot}
\caption[he]{\footnotesize \narrower
Finite-size scaling plot for the interface free energy 
of the three dimensional Heisenberg model.}
\label{scaling.plot.n3}
\end{figure}
 
The data collapse for the $n=3$ case onto the function $\Sigma$ as 
determined by the least-squares fit is shown in 
Figure~\ref{scaling.plot.n3}. 

Finally we remark that, although in each of the cases $n=1$, 2 and 3
the finite-size effect appears to be small for $L\geq6$, it is large
for $L=4$. For this reason the $L=4$ data were not included in the
fits.

\acknowledgments
This research was supported  by the (US) National Science Foundation
through Grant \# DMR-9214669, by the Office of Naval Research and by
the NATO through Grant \# CRG 910152.  This research was conducted in
part using the resources of the Cornell Theory Center, which receives
major funding from the National Science Foundation (NSF) and New York
State, with additional support from the Advanced Research Projects
Agency (ARPA), the National Center for Research Resources at the
National Institutes of Health (NIH), IBM Corporation, and other members
of the center's Corporate Research Institute.

\appendix
\section{Scaling plot Parameter estimates}

Tables \ref{tab.Isingfit} through \ref{tab.Heisfit} contain estimates
of the parameters used in the finite-size scaling plots for the
interface free energy of O($n$) models, as discussed in
Section~\ref{sec.applic}.

\begin{table}
\caption[Isingfit]{\footnotesize \narrower
Parameters, as defined in the text, and their standard errors for
the scaling function of the interfacial free energy of the 
three-dimensional Ising model.}
\label{tab.Isingfit}
\bigskip
\begin{tabular}{|l|c|c|}
                       &$n \ge 6$&                   $n \ge 8$\\
\hline
$K_{\/\rm c}$&  0.22162  $\pm$  0.00002  &   0.22165  $\pm$  0.00003  \\
$y_{\/\rm T}$&  1.583    $\pm$  0.004    &   1.594    $\pm$  0.009    \\
$\Sigma_0$   &  0.6171   $\pm$  0.0007   &   0.6194   $\pm$  0.0025   \\
$\Sigma_1$   &  2.6111   $\pm$  0.0176   &   2.5650   $\pm$  0.0505   \\
$\Sigma_2$   &  6.0475   $\pm$  0.1001   &   5.8047   $\pm$  0.2565   \\
$\Sigma_3$   &  9.2362   $\pm$  0.3052   &   8.4073   $\pm$  0.6724   \\
$\Sigma_4$   &  6.0087   $\pm$  0.6350   &   4.8601   $\pm$  1.0249   \\
$\Sigma_5$   &-13.6165   $\pm$  0.9830   & -10.8331   $\pm$  1.7414   \\
$\Sigma_6$   &-33.2578   $\pm$  4.0574   & -24.7108   $\pm$  5.6481   \\
$\Sigma_7$   &  4.4790   $\pm$  3.3849   &   1.4272   $\pm$  4.2080   \\
$\Sigma_8$   & 70.6918   $\pm$ 11.4018   &  46.6641   $\pm$ 14.0107   \\
$\Sigma_9$   & 28.0314   $\pm$  7.9088   &  21.4636   $\pm$  8.5327   \\
$\Sigma_{10}$&-69.1548   $\pm$ 14.3387   & -40.6066   $\pm$ 15.7549   \\
$\Sigma_{11}$&-43.4754   $\pm$ 10.0932   & -27.7038   $\pm$ 10.3867   \\
$\Sigma_{12}$& 25.2137   $\pm$  6.4840   &  13.2245   $\pm$  6.4356   \\
$\Sigma_{13}$& 18.8658   $\pm$  4.8588   &  10.5010   $\pm$  4.6564   \\
$a$          & -2.65     $\pm$  0.16     &  -2.65     $\pm$  0.37     \\
\end{tabular}
\end{table}
\begin{table}
\caption[XYfit]{\footnotesize \narrower
Parameters, as defined in the text, and their standard errors for
the scaling function of the interfacial free energy of the 
three-dimensional planar model.}
\label{tab.XYfit}
\bigskip
\begin{tabular}{|l|c|c|}
                       &$n \ge 6$&                   $n \ge 8$\\
\hline
$K_{\/\rm c}$& 0.45410   $\pm$ 0.00003  &  0.45413   $\pm$ 0.00005  \\
$y_{\/\rm T}$& 1.491     $\pm$ 0.003    &  1.487     $\pm$ 0.006    \\
$\Sigma_0$   & 1.2448    $\pm$ 0.0010   &  1.2469    $\pm$ 0.0033   \\
$\Sigma_1$   & 2.5592    $\pm$ 0.0144   &  2.5929    $\pm$ 0.0345   \\
$\Sigma_2$   & 2.4285    $\pm$ 0.0439   &  2.4738    $\pm$ 0.0796   \\
$\Sigma_3$   & 0.9881    $\pm$ 0.0623   &  0.9544    $\pm$ 0.1031   \\
$\Sigma_4$   &-0.5096    $\pm$ 0.0664   & -0.4292    $\pm$ 0.1050   \\
$\Sigma_5$   &-0.7770    $\pm$ 0.1579   & -0.5171    $\pm$ 0.2741   \\
$\Sigma_6$   & 0.1737    $\pm$ 0.0754   &  0.0793    $\pm$ 0.1204   \\
$\Sigma_7$   & 0.4346    $\pm$ 0.1503   &  0.1847    $\pm$ 0.2567   \\
$a$          &-0.7805    $\pm$ 0.1159   & -0.9245    $\pm$ 0.2259   \\
\end{tabular}
\end{table}
\begin{table}
\caption[Heisfit]{\footnotesize \narrower
Parameters, as defined in the text, and their standard errors for
the scaling function of the interfacial free energy of the 
three-dimensional Heisenberg model. The Monte Carlo data were
taken relatively close to $K_{\rm c}$, so that the temperature
exponent  $y_{\rm T}$ is not accurately determined. The accuracy 
of $K_{\rm c}$ is unaffected.}
\label{tab.Heisfit}
\bigskip
\begin{tabular}{|l|c|c|}
                       &$n \ge 6$&                   $n \ge 8$\\
\hline
$K_{\/\rm c}$& 0.69291  $\pm$  0.00004 & 0.69294  $\pm$  0.00008 \\
$y_{\/\rm T}$& 1.44     $\pm$  0.07    & 1.55     $\pm$  0.18    \\
$\Sigma_0$   & 1.8919   $\pm$  0.0015  & 1.8933   $\pm$  0.0043  \\
$\Sigma_1$   & 2.4563   $\pm$  0.3123  & 1.9036   $\pm$  0.7665  \\
$\Sigma_2$   & 0.7991   $\pm$  0.3005  & 0.5097   $\pm$  0.4651  \\
\end{tabular}
\end{table}


\begin{thebibliography}{99}
\bibitem{KraWa} H.A. Kramers and G.H. Wannier, Phys. Rev.
{\bf 60}, 252 (1941).
\bibitem{Trot} H.F. Trotter, Proc. Am. Math. Soc. 
{\bf 10}, 545 (1959).
\bibitem{Runge.92.b}  For a discussion of these correlations see
K.J. Runge, \prb {\bf 45}, 12 292 (1992).
\bibitem{NighGraKos95}
M.P. Nightingale, E. Granato and J.M. Kosterlitz,
\prb {\bf 52}, 7402 (1995).
\bibitem{NB.PRL.60} M.P. Nightingale and H.W.J. Bl\"ote, 
\prl {\bf 60}, 1662 (1988).
\bibitem{TMreview}M.P. Nightingale,  in {\it Finite-size scaling and
simulation of statistical mechanical systems,} V. Privman, ed. (World
Scientific, Singapore 1990), p.287-351.
\bibitem{UNR} C.J. Umrigar, M.P. Nightingale, and K.J. Runge, 
J. Chem. Phys.  {\bf 99}, 2865 (1993).
\bibitem{Hetherington}J.H. Hetherington, \pra {\bf 30}, 2713 (1984).
\bibitem{NB.PRB.33} M.P. Nightingale and H.W.J. Bl\"ote, 
\prb {\bf 33}, 659 (1986).
\bibitem{UmrigarCeperleyPriv}D.M. Ceperley and C.J. Umrigar, private
communication.
\bibitem{Kalospri}M.H. Kalos, J. Comp. Phys. {\bf 1}, 257 (1966);
the original idea of "forward walking" predates this paper
(M.H. Kalos private communication).  For further references see Ref. 11
of Ref.\onlinecite{Runge.92.a}.
\bibitem{Runge.92.a}  K.J. Runge, \prb {\bf 45}, 7229 (1992).
\bibitem{ck.79}D.M. Ceperley and M.H. Kalos, {\it Monte Carlo
Methods in Statistical Physics}, edited by K. Binder (Springer, Berlin,
1979).
\bibitem{UWW} C.J. Umrigar, K.G. Wilson and J.W. Wilkins,  
\prl {\bf 60}, 1719 (1988);
{\it Computer Simulation Studies in Condensed Matter
Physics}, edited by D.P. Landau, K.K. Mon, and H.-B. Sch\"uttler, {\it
Springer Proceedings in Physics}\ 33 (Springer-Verlag, Berlin, 1988)
p.185.
\bibitem{GraNigh93} E. Granato and M.P. Nightingale,
\prb {\bf 48}, 7438 (1993).
\bibitem{BLH} H.W.J. Bl\"ote, E. Luijten and J.R. Heringa,
J. Phys. A {\bf 28}, 6289 (1995).
\bibitem{BHHSM}  
H.W.J. Bl\"ote, J.R. Heringa, A. Hoogland, E.W. Meyer and T.S. Smit,
preprint (1996).
\bibitem{GT}
R. Gupta and P. Tamayo, preprint; to appear in Int. J. Mod. Phys (1996).
\bibitem{FerLan} A.M. Ferrenberg and D.P. Landau, 
\prb {\bf 44}, 5081 (1991).
\bibitem{ShawFisher} L. Shaw and M.E. Fisher, \pra 39, 2189 (1989).
\bibitem{KKMonPRL88} K.K. Mon, \prl {\bf 60}, 2749 (1988).
\bibitem{Ferer} M. Ferer, M.A. Moore and M. Wortis,
\prb {\bf 8}, 5205 (1973).
\bibitem{Ohno} K. Ohno, Y. Okabe and A. Morita, Prog. Theor.
Phys. {\bf 71}, 714 (1984).
\bibitem{LanPanB} D.P. Landau, R. Pandey and K. Binder,
\prb {\bf 39}, 12302 (1989).
\bibitem{LGZJ}
 J.C. Le Guillou and J. Zinn-Justin, \prb {\bf 21} 3976 (1980)
\bibitem{RitFis} D.S Ritchie and M.E. Fisher, 
\prb 5, 2668, 1972.
\bibitem{ChenFerLan} K. Chen, A.M. Ferrenberg and D.P. Landau, 
J. Appl. Phys. {\bf 73}, 5488 (1993).
\end{thebibliography}
\end{document}